\documentclass[fleqn,usenatbib]{mnras}

\usepackage{newtxtext,newtxmath}

\usepackage[T1]{fontenc}

\DeclareRobustCommand{\VAN}[3]{#2}
\let\VANthebibliography\thebibliography
\def\thebibliography{\DeclareRobustCommand{\VAN}[3]{##3}\VANthebibliography}

\usepackage{graphicx}	
\usepackage{amsmath}	
\usepackage{subcaption}

\title[AT2021lwx as an extraordinary accretion event]
{Multiwavelength observations of the extraordinary accretion event AT2021lwx}

\author[P. Wiseman et al.]{
\parbox{\textwidth}{
\Large
P. Wiseman,$^{1}$\thanks{E-mail: p.s.wiseman@soton.ac.uk (PW)}
Y. Wang,$^{1,2}$
S. H\"onig,$^{1}$
N. Castro-Segura,$^{1}$
P. Clark,$^{3}$ 
C. Frohmaier,$^{1}$ 
M. D. Fulton,$^{4}$ 
G. Leloudas,$^{5}$
M. Middleton,$^{1}$
T. E. M\"uller-Bravo,$^{6,7}$ 
A. Mummery,$^{8}$
M. Pursiainen,$^{5}$ 
S. J. Smartt,$^{8,4}$
K. Smith,$^{4}$ 
M. Sullivan,$^{1}$,
J.~P. Anderson,$^{9,10}$
J. A. Acosta Pulido,$^{11,12}$ 
P. Charalampopoulos,$^{13}$ 
M. Banerji,$^{1}$
M. Dennefeld,$^{14}$
L. Galbany,$^{6,7}$
M. Gromadzki,$^{15}$ 
C. P. Guti\'errez$^{16,17}$
N. Ihanec,$^{15}$
E. Kankare,$^{13,18}$ 
A. Lawrence,$^{19}$ 
B. Mockler,$^{20}$
T. Moore,$^{4}$ 
M. Nicholl,$^{4}$ 
F. Onori,$^{21}$ 
T. Petrushevska,$^{22}$
F. Ragosta,$^{23}$ 
S. Rest,$^{24}$ 
M. Smith,$^{25}$
T. Wevers,$^{9}$ 
R. Carini,$^{23}$ 
T.-W. Chen,$^{26,27}$
K. Chambers,$^{28}$ 
H. Gao,$^{28}$ 
M. Huber,$^{28}$
C. Inserra,$^{29}$
E. Magnier,$^{28}$
L. Makrygianni,$^{30}$
M. Toy,$^{1}$ 
F. Vincentelli,$^{10,11}$
D. R. Young,$^{4}$ 
}\vspace{0.04in}\\
$^{1}$School of Physics and Astronomy, University of Southampton, Southampton, SO17 1BJ, UK\\
$^{2}$Key Laboratory of Optical Astronomy, National Astronomical Observatories, Chinese Academy of Sciences, Beijing 100101, China\\
$^{3}$Institute of Cosmology and Gravitation, University of Portsmouth, Portsmouth, PO1 3FX, UK\\
$^{4}$Astrophysics Research Centre, School of Mathematics and Physics, Queen's University Belfast, BT7 1NN, UK\\
$^{5}$DTU Space, National Space Institute, Technical University of Denmark, Elektrovej 327, 2800 Kgs. Lyngby, Denmark \\
$^{6}$Institute of Space Sciences (ICE, CSIC), Campus UAB, Carrer de Can Magrans, s/n, E-08193 Barcelona, Spain\\
$^{7}$Institut d’Estudis Espacials de Catalunya (IEEC), E-08034 Barcelona, Spain\\
$^{8}$Department of Physics, University of Oxford, Denys Wilkinson Building, Keble Road, Oxford OX1 3RH, UK\\
$^{9}$European Southern Observatory, Alonso de C\'ordova 3107, Casilla 19, Santiago, Chile\\
$^{10}$Millennium Institute of Astrophysics MAS, Nuncio Monsenor Sotero Sanz 100, Off. 104, Providencia, Santiago, Chile\\
$^{11}$Instituto de Astrofísica de Canarias (IAC), E-38200 La Laguna, Tenerife, Spain\\
$^{12}$Universidad de La Laguna (ULL), Departamento de Astrofisica, E-38206, Tenerife, Spain\\
$^{13}$Department of Physics and Astronomy, University of Turku, FI-20014 Turku, Finland\\
$^{14}$  Institut d’Astrophysique de Paris (IAP), CNRS \& Sorbonne Université, 75014 Paris, France\\
$^{15}$Astronomical Observatory, University of Warsaw, Al. Ujazdowskie 4, 00-478 Warszawa, Poland\\
$^{16}$Finnish Centre for Astronomy with ESO (FINCA), FI-20014 University of Turku, Finland\\
$^{17}$Tuorla Observatory, Department of Physics and Astronomy, FI-20014 University of Turku, Finland\\
$^{18}$Turku Collegium for Science, Medicine and Technology, University of Turku, FI-20014 Turku, Finland\\
$^{19}$Institute for Astronomy, University of Edinburgh, Royal Observatory, Blackford Hill, Edinburgh EH9 3HJ\\
$^{20}$Department of Astronomy and Astrophysics, University of California, Santa Cruz, Santa Cruz, CA 95064, USA \\
$^{21}$INAF - Osservatorio Astronomico d'Abruzzo, Via M. Maggini snc, I-64100 Teramo, Italy\\
$^{22}$ University of Nova Gorica, Center for Astrophysics and Cosmology, Vipavska 11c, SI-5270 Ajdovščina, Slovenia\\
$^{23}$ INAF-Osservatorio Astronomico di Roma, via Frascati 33, 00078 Monte Porzio Catone (RM), Italy\\ 
$^{24}$Department of Physics and Astronomy, The Johns Hopkins University, Baltimore, MD 21218, USA\\
$^{25}$Univ Lyon, Univ Claude Bernard Lyon 1, CNRS, IP2I Lyon / IN2P3, IMR 5822, F-69622 Villeurbanne, France\\
$^{26}$Technische Universit{\"a}t M{\"u}nchen, TUM School of Natural Sciences, Physik-Department, James-Franck-Stra{\ss}e 1, 85748 Garching, Germany\\
$^{27}$Max-Planck-Institut f{\"u}r Astrophysik, Karl-Schwarzschild Stra{\ss}e 1, 85748 Garching, Germany\\
$^{28}$Institute for Astronomy, University of Hawaii, 2680 Woodlawn Drive, Honolulu, HI 96822, USA\\
$^{29}$Cardiff Hub for Astrophysics Research and Technology, School of Physics \& Astronomy, Cardiff University, Queens Buildings, The Parade, Cardiff, CF24 3AA, UK\\
$^{30}$The School of Physics and Astronomy, Tel Aviv University, Tel Aviv 69978, Israel\\
}

\date{Accepted XXX. Received YYY; in original form ZZZ}

\pubyear{2023}

\begin{document}
\label{firstpage}
\pagerange{\pageref{firstpage}--\pageref{lastpage}}
\maketitle

\begin{abstract}
 We present observations from X-ray to mid-infrared wavelengths of the most energetic non-quasar transient ever observed, AT2021lwx. Our data show a single optical brightening by a factor $>100$ to a luminosity of $7\times10^{45}$\,erg\,s$^{-1}$, and a total radiated energy of $1.5\times10^{53}$\,erg, both greater than any known optical transient. The decline is smooth and exponential and the ultra-violet - optical spectral energy distribution resembles a black body with temperature $1.2\times10^4$\,K. Tentative X-ray detections indicate a secondary mode of emission, while a delayed mid-infrared flare points to the presence of dust surrounding the transient. The spectra are similar to recently discovered optical flares in known active galactic nuclei but lack some characteristic features. The lack of emission for the previous seven years is inconsistent with the short-term, stochastic variability observed in quasars, while the extreme luminosity and long timescale of the transient disfavour the disruption of a single solar-mass star. The luminosity could be generated by the disruption of a much more massive star, but the likelihood of such an event occurring is small. A plausible scenario is the accretion of a giant molecular cloud by a dormant black hole of $10^8 - 10^9$ solar masses. AT2021lwx thus represents an extreme extension of the known scenarios of black hole accretion.
\end{abstract}

\begin{keywords}
accretion, accretion discs -- transients:  tidal disruption events -- Transients -- galaxies: active -- quasars: supermassive black holes
\end{keywords}



\section{Introduction}

The accretion of matter onto supermassive black holes (SMBHs) is the most efficient known process of extracting energy in the Universe, with general relativistic effects combined with complex magnetohydrodynamics and highly non-thermal radiative processes causing spectacular electromagnetic phenomena. This accretion can happen at a widely varying range of rates: steadily-accreting SMBHs, called active galactic nuclei (AGN), are the central engines of galaxies, which accrete gas for Myr at a time with a rate that typically varies on timescales of seconds to years \citep[e.g.][]{McHardy2006,Scaringi2015}.

As opposed to typical variability of AGN caused by stochastic accretion, tidal disruption events (TDEs; \citealt{Hills1975,Rees1977}) represent a much shorter, yet far more violent accretion episode. These phenomena are caused by the destruction of a star in the vicinity of a SMBH (or intermediate-mass black hole; \citealt{Angus2022,Yao2023}) due to tidal forces. TDEs usually reveal themselves through a single flare observed in optical/ultraviolet (UV)/X-rays with a smooth rise and decaying light curve. Not all SMBHs are equally likely to produce a TDE: for SMBHs with masses $M_\mathrm{BH}\gtrsim10^{8}\,\mathrm{M_\odot}$, the tidal radius of a solar mass star lies within the innermost stable circular orbit (ISCO) of the black hole, which suppresses or completely prohibits TDEs (the exact limit depends on stellar type and SMBH spin). Indeed, observed or model-inferred masses of TDEs suggest $M_\mathrm{BH} \sim 5\times10^{5}-10^{7}\,\mathrm{M_\odot}$ and $M_* \sim 0.6-13\,\mathrm{M_\odot}$ \citep{Mockler2019,Ryu2020}, though the high end of this stellar mass range is highly uncertain. TDEs display broad emission lines, often of hydrogen and/or helium, and many show ionised iron and Bowen flourescence lines of nitrogen and oxygen \citep[e.g.][]{Arcavi2014,Leloudas2019,VanVelzen2021}, indicative of the presence of a UV-bright accretion disk.

The clear differences between regular AGN variability and TDEs are muddied by the addition of recently-discovered changing look AGN (CLAGNs; \citealt{LaMassa2015}) and SMBH `re-ignitions' \citep{Trakhtenbrot2019}.  CLAGNs tend to exhibit changes in their spectral properties such as line widths and continuum slopes \citep{Komossa2022}, sometimes accompanied by X-ray outbursts and a change in optical brightness but critically while maintaining their usual variable optical light curve. On the other hand, re-ignitions and flares appear as a single highly luminous optical transient \citep{Kankare2017,Gromadzki2019,Trakhtenbrot2019,Frederick2021}. The spectra of these flares, which are often found in narrow-line Seyfert 1 (NLSy1) host galaxies, show ubiquitous hydrogen Balmer emission lines as well as many of the same helium and Bowen features of TDEs \citep{Frederick2021}, although the line profiles (particularly of hydrogen) tend to be narrower. While the high luminosity of the flares necessitates a large amount of material being accreted, the exact nature remains a mystery, with various sources labelled as possible abnormal TDEs or sudden changes in the accretion flow of already-accreting SMBHs.

Here we present multiwavelength observations of AT2021lwx, the most luminous of such flares ever detected. AT2021lwx, at $z=0.9945$, displays properties spanning TDEs, SMBH flares, and AGN, while unlike all of those classes there is no host galaxy visible to deep limits in pre-outburst survey data. Our observations (Section \ref{sec:obs}) span nearly three years and from MIR to X-ray wavelengths. We present details of our photometric (Section \ref{sec:phot}) and spectroscopic (Section \ref{sec:spec}) modelling, motivating our discussion of the potential scenarios of the event (Section \ref{sec:disc}), after which we conclude by interpreting AT2021lwx as an accretion of a non-stellar, gaseous object onto a large SMBH.

Throughout the paper, we use `days' to refer to the observer frame and `d' for the rest frame at $z=0.9945$. All photometry is corrected for Galactic foreground extinction according to \citet{Schlafly2011} assuming a \citet{Fitzpatrick1999} extinction curve. Magnitudes are presented in the AB system \citep{Oke1983}. Upper limits are reported at the $3\sigma$ level, while uncertainties are $1\sigma$ Where necessary we assume a spatially flat lambda cold dark matter cosmological model with $H_0 = 70$~km~s$^{-1}$~Mpc$^{-1}$ and $\Omega_M=0.3$.

\begin{table}
 \caption{Summary of the multiwavelength observations of AT2021lwx.}
 \label{tab:example}
 \begin{tabular}{llll}
  \hline
  MJD & Epoch & Instrument & Bands \\
     & (rest frame d) & &\\
  \hline
 & & Photometry  &\\
  \hline
  58750 to 59937 & $-271$ to $+323$ & ZTF & $g$, $r$\\
  58113 to 59908 & $-590$ to $+309$ & ATLAS & c, o\\
  56803 to 59885 , & $-1247$ to $+298$, & WISE & $W1$, $W2$\\
  59923, 59966 & $+317$, $+338$ &\textit{Swift} UVOT & $uvw2, uvm2,$\\
  & & & $ uvw1, u, b, v$  \\
  \hline
  \hline
  & &X-ray  & \\
  \hline
59923, 59966 & $+317$, $+338$ &\textit{Swift} XRT & $0.3-10$\, keV \\
  \hline
  \hline
 & &  Spectroscopy & \\
  \hline
  59345, 59844 & $+27$, $+277$ & NTT EFOSC2 & $3685 - 9315$\,\AA\\
  59931 & $+320$ & GTC EMIR & $0.85 - 1.35$\,$\mu$m\\
   59949 & $+329$ & GTC EMIR & $1.45 - 2.42$\,$\mu$m\\
 \end{tabular}
\end{table}


\begin{figure*}
\centering
\begin{subfigure}[][][t]{\textwidth}
    \includegraphics[width=\textwidth]{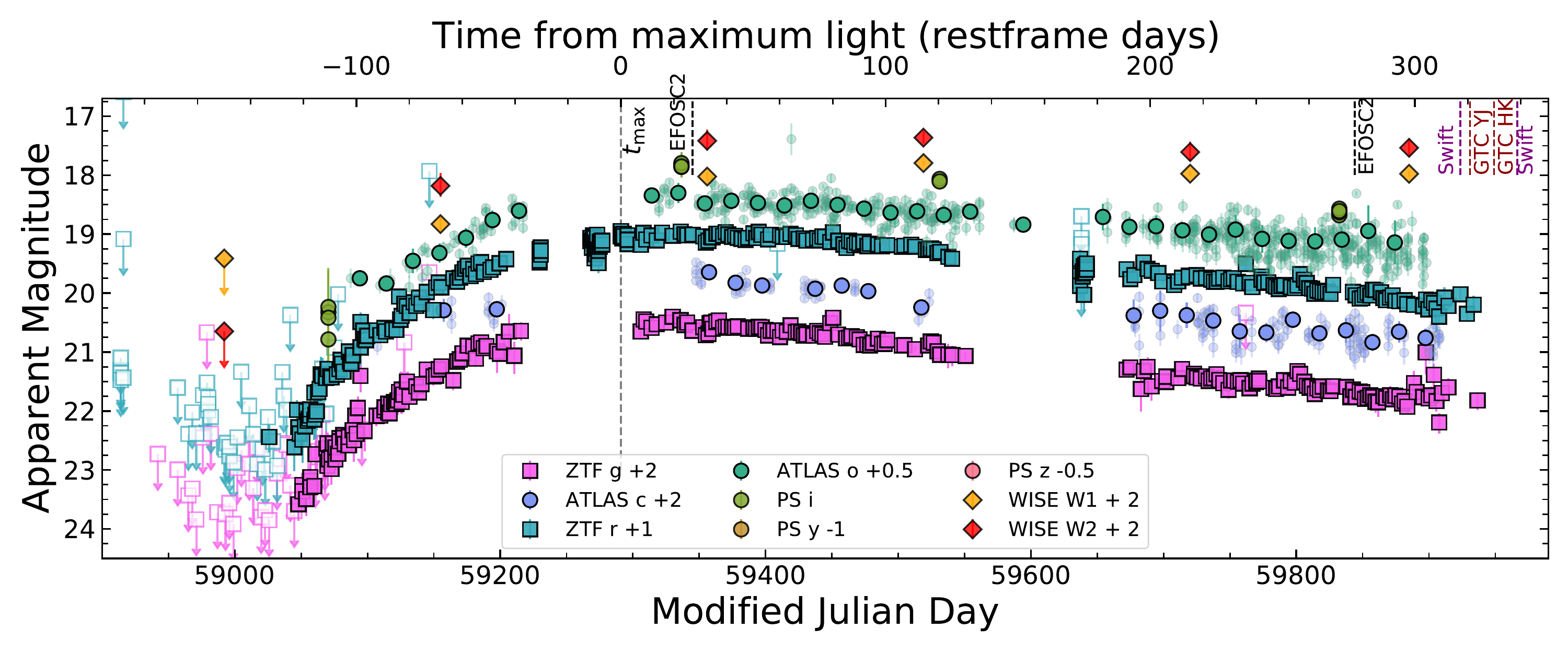}
    \label{fig:lc_all}
\end{subfigure}
\hfill
\begin{subfigure}[][][b]{0.49\textwidth}
    \includegraphics[width=\textwidth]{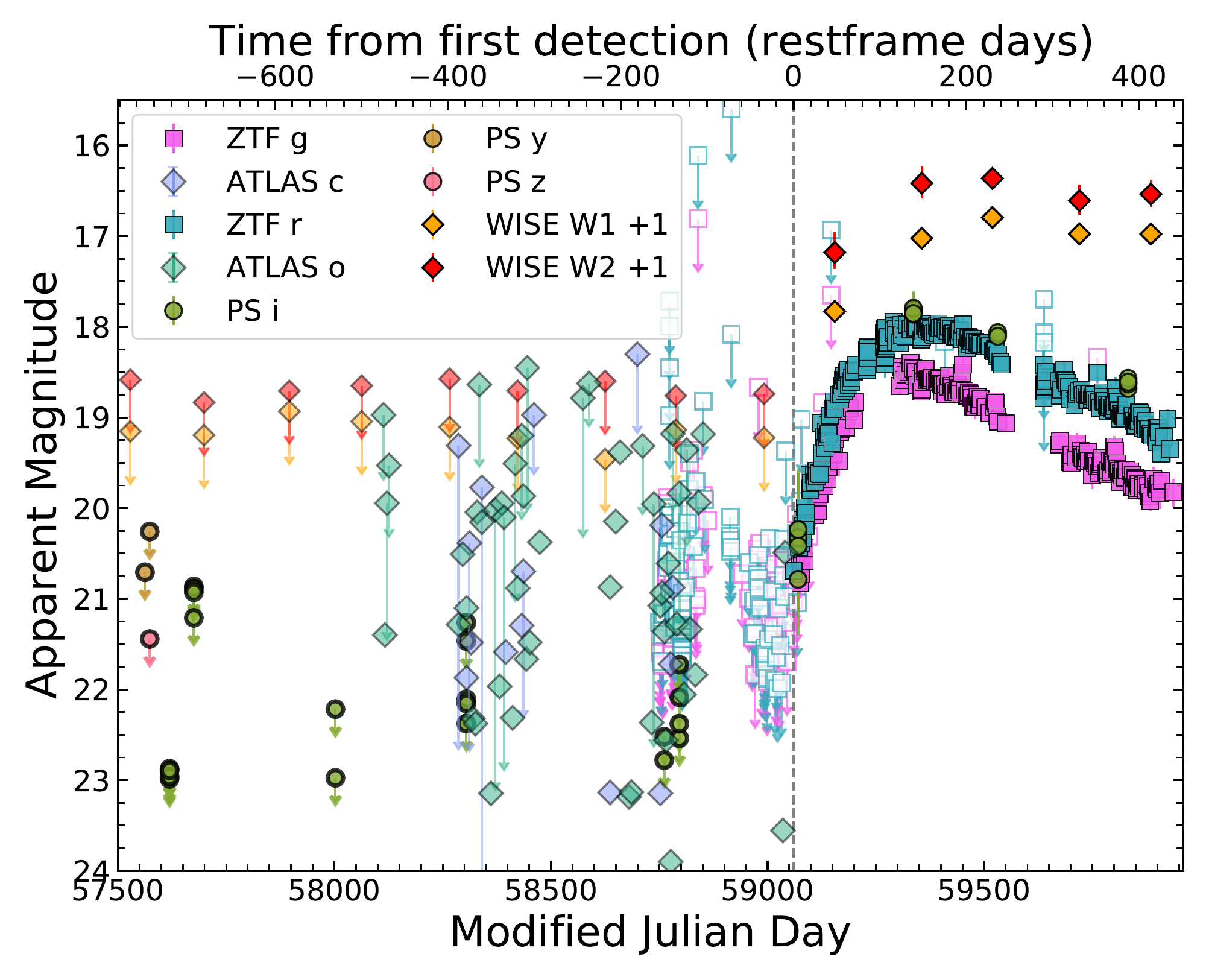}
    \label{fig:lc_lims}
\end{subfigure}
\hfill
\begin{subfigure}[][][b]{0.49\textwidth}
    \includegraphics[width=\textwidth]{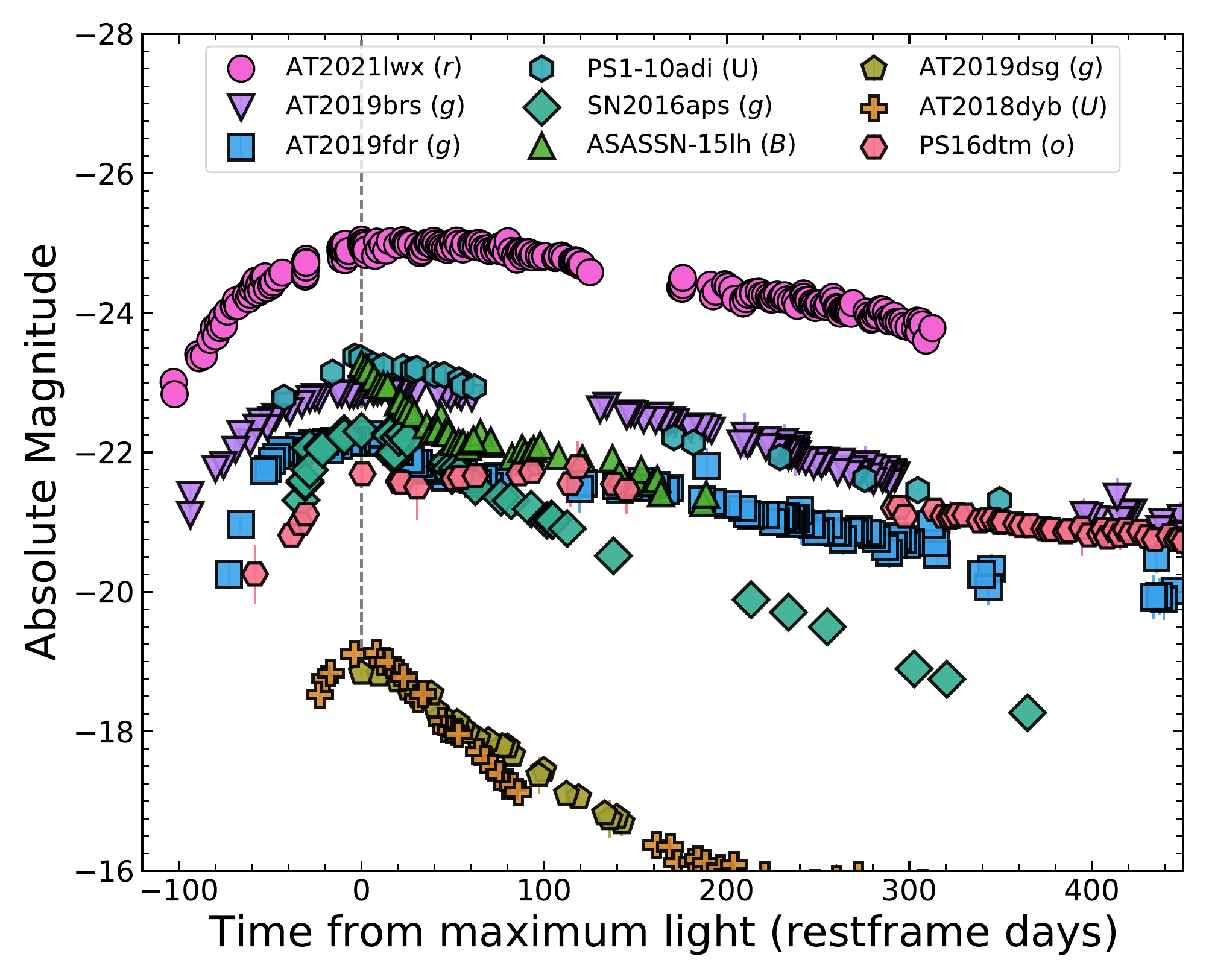}
    \label{fig:lc_comp}
\end{subfigure}
        
\caption{upper) Light curve of AT2021lwx. Epochs of our multiwavelength follow-up observations are marked with dashed lines; lower left) Pan-STARRS upper limits up to 750\,d (rest frame) before the first detection of AT2021lwx; lower right) Comparison to similar transients: NLSy1 accretion events  AT2019fdr and AT20219brs \citep{Frederick2021} and the prototype nuclear flare PS1-10adi \citep{Kankare2017}; the most luminous known likely TDE ASASSN-15lh \citep{Leloudas2016}; the most luminous known supernova SN2016aps \citep{Nicholl2020a}; a (possibly jetted) TDE AT2019dsg \citep{VanVelzen2021}, a typical TDE AT2018dyb \citep{Leloudas2019}, and the MIR-brightening TDE PS16dtm \citep{Petrushevska2023}.}
\label{fig:lc_main}
\end{figure*}

\begin{figure}%
\centering
\includegraphics[width=0.49\textwidth]{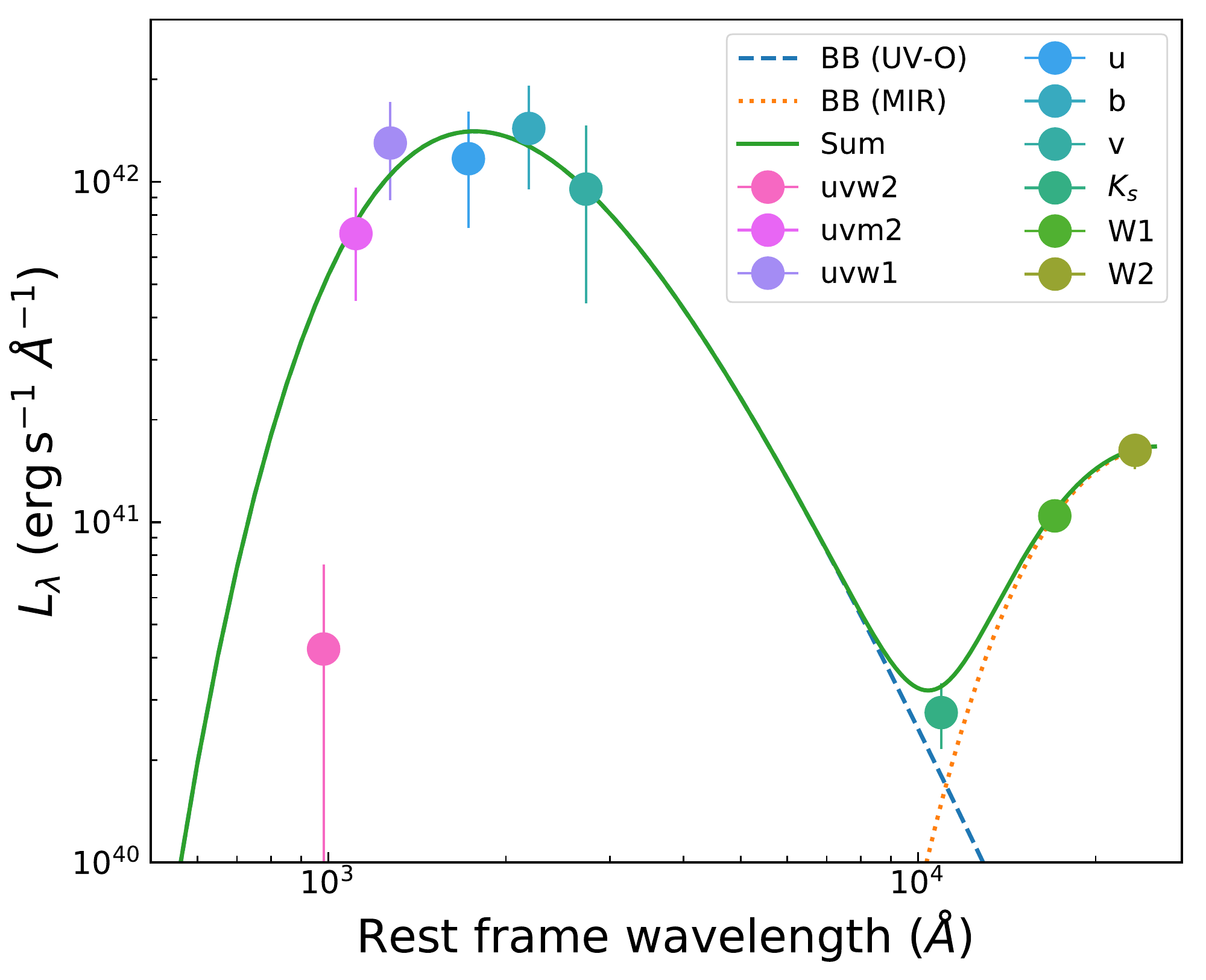}
\caption{The rest frame UV to MIR SED of AT2021lwx at $+338$\,d. The MIR data come from $+114$\,d and are not scaled as we do not predict their evolution. Data are corrected for MW reddening, but not for any host extinction. The uvw2 point is not included in the black body fit due to likely Lyman-$\alpha$ absorption.}\label{fig:SED_all}
\end{figure}

\begin{figure*}%
\centering
\includegraphics[width=\textwidth]{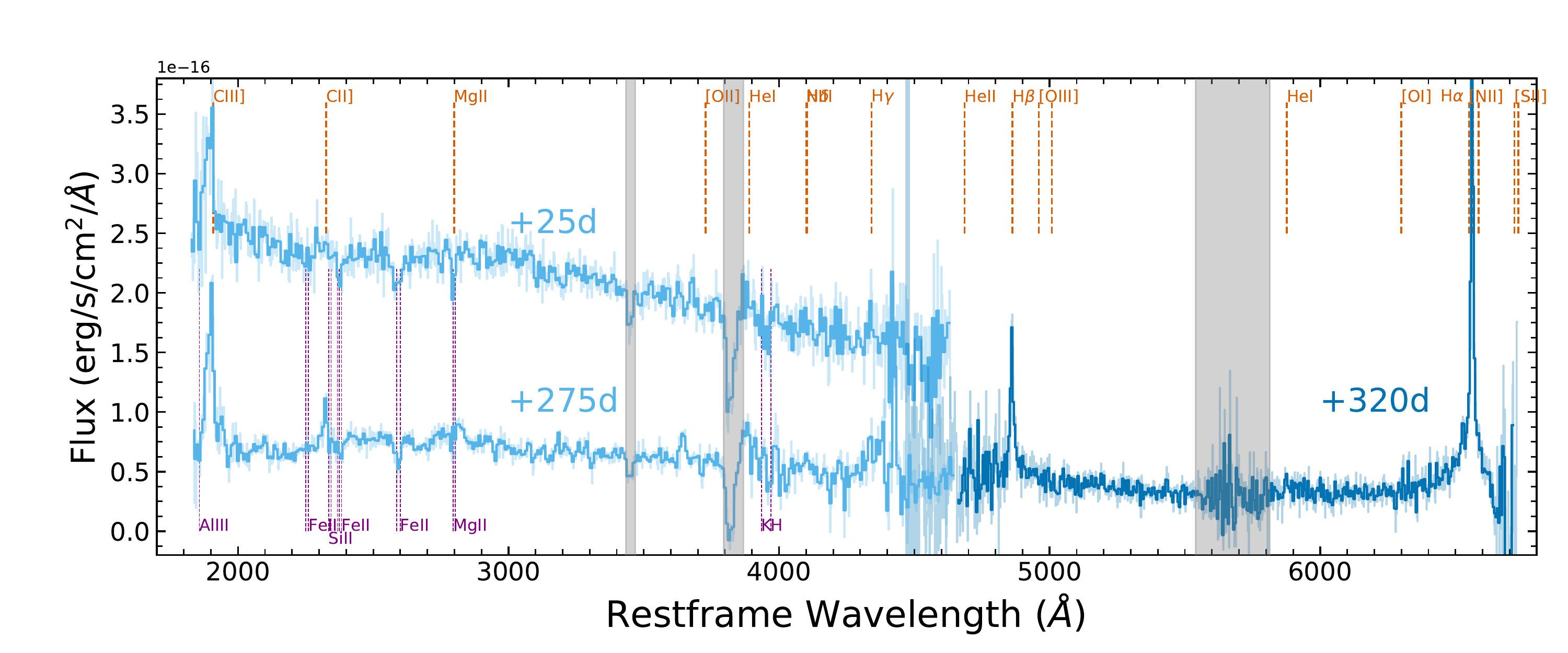}
\caption{Rest frame UV and optical spectra of AT2021lwx. For clarity, the GTC EMIR +320\,d spectrum (dark blue) has been scaled to match the optical NTT EFOSC2 spectrum from +275\,d. Common emission features are shown in gold while absorption features are marked in purple. Telluric absorption has been masked in grey.}\label{fig:our_spec}
\end{figure*}

\begin{figure*}%
\centering
\includegraphics[width=\textwidth]{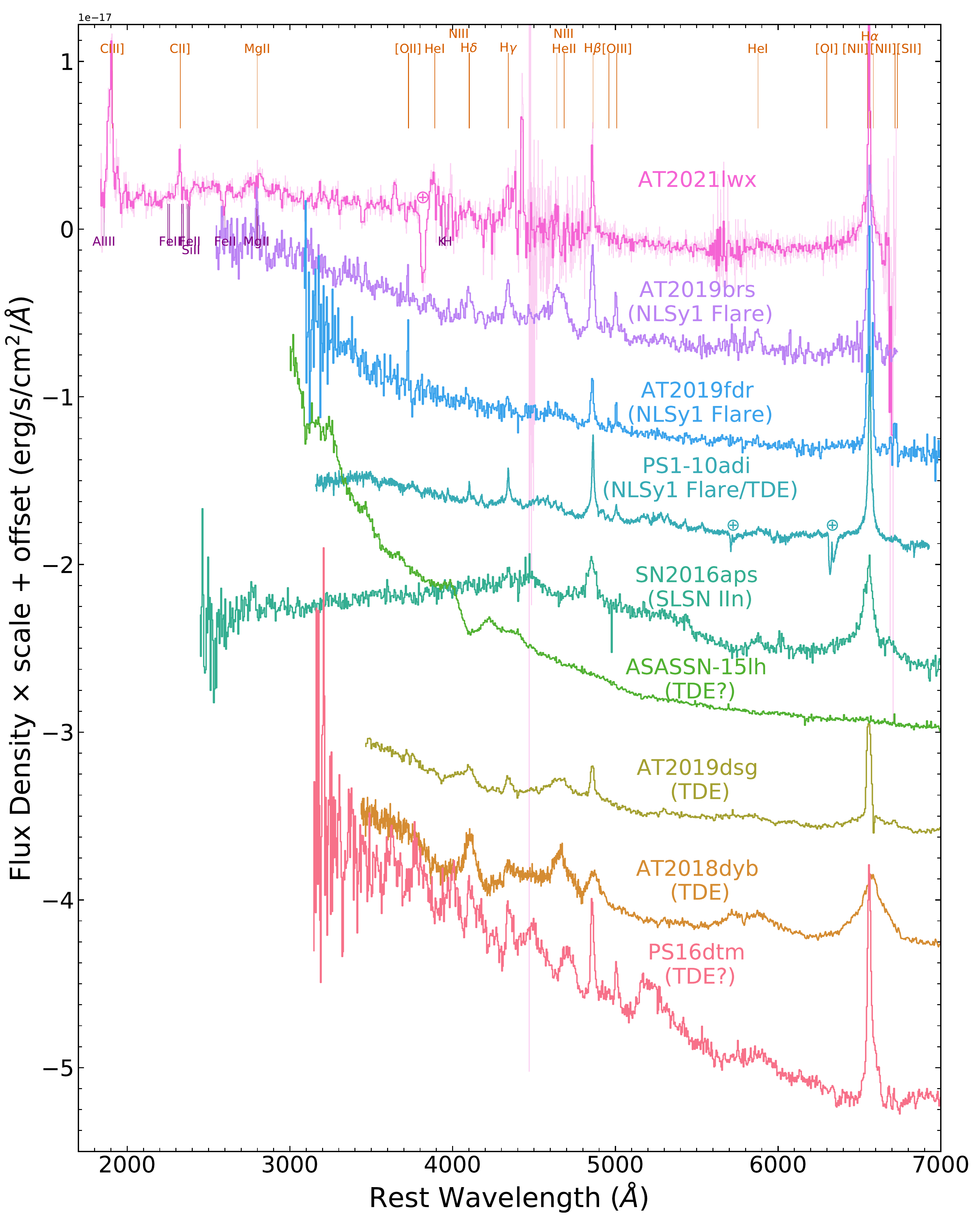}
\caption{Rest frame UV and optical spectra of AT2021lwx as well as the same objects compared in Fig. \ref{fig:lc_main}. Spectra have been shifted and scaled in order to provide a qualitative comparison. Telluric features are marked with $\oplus$. Spectra have been retrieved from the WISeREP spectral database\citep{Yaron2012}.}\label{fig:comparison_spec}
\end{figure*}


\begin{figure*}
\centering
\begin{subfigure}[][][t]{0.49\textwidth}
    \includegraphics[width=\textwidth]{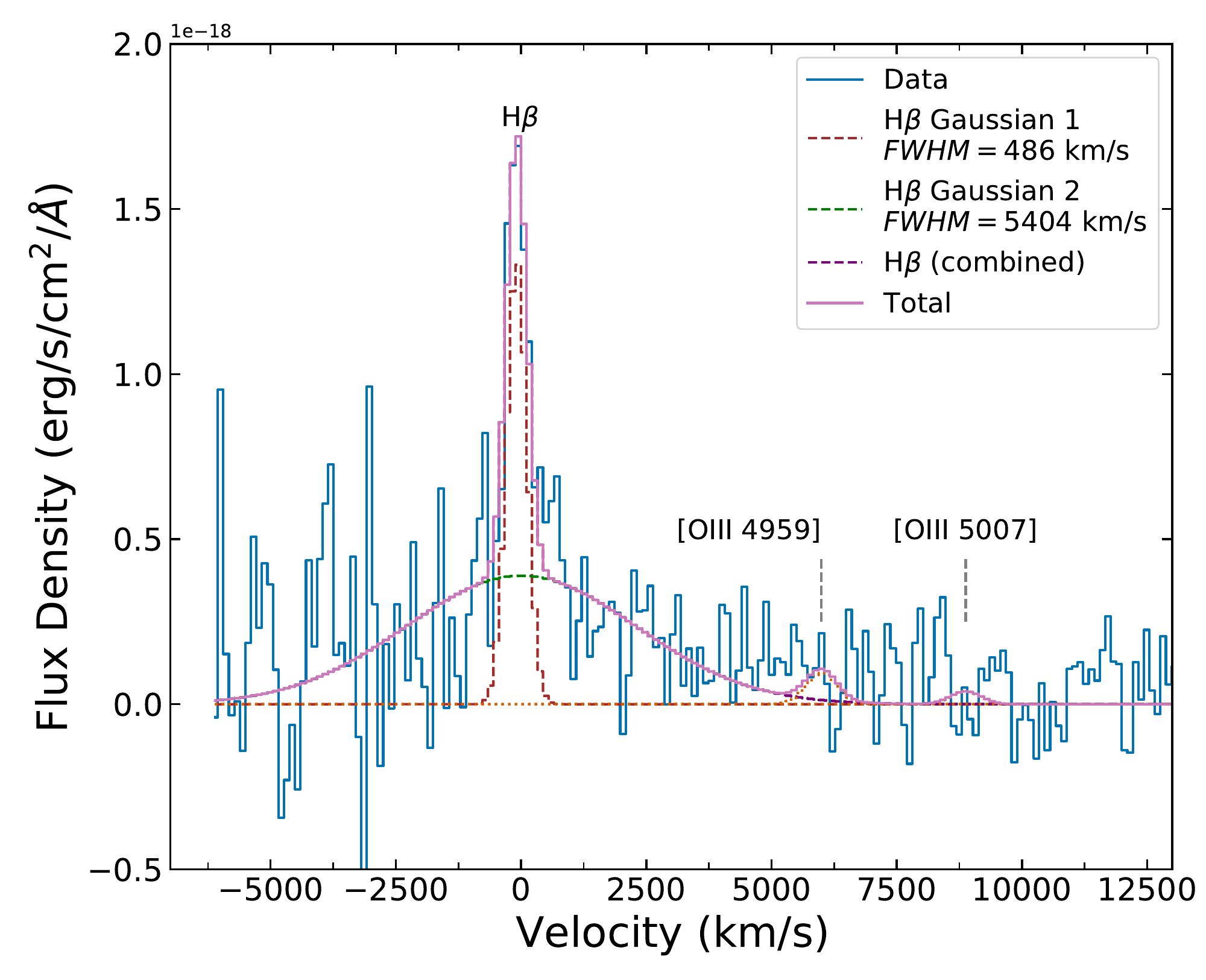}
    \label{fig:hbeta}
\end{subfigure}
\hfill
\begin{subfigure}[][][b]{0.49\textwidth}
    \includegraphics[width=\textwidth]{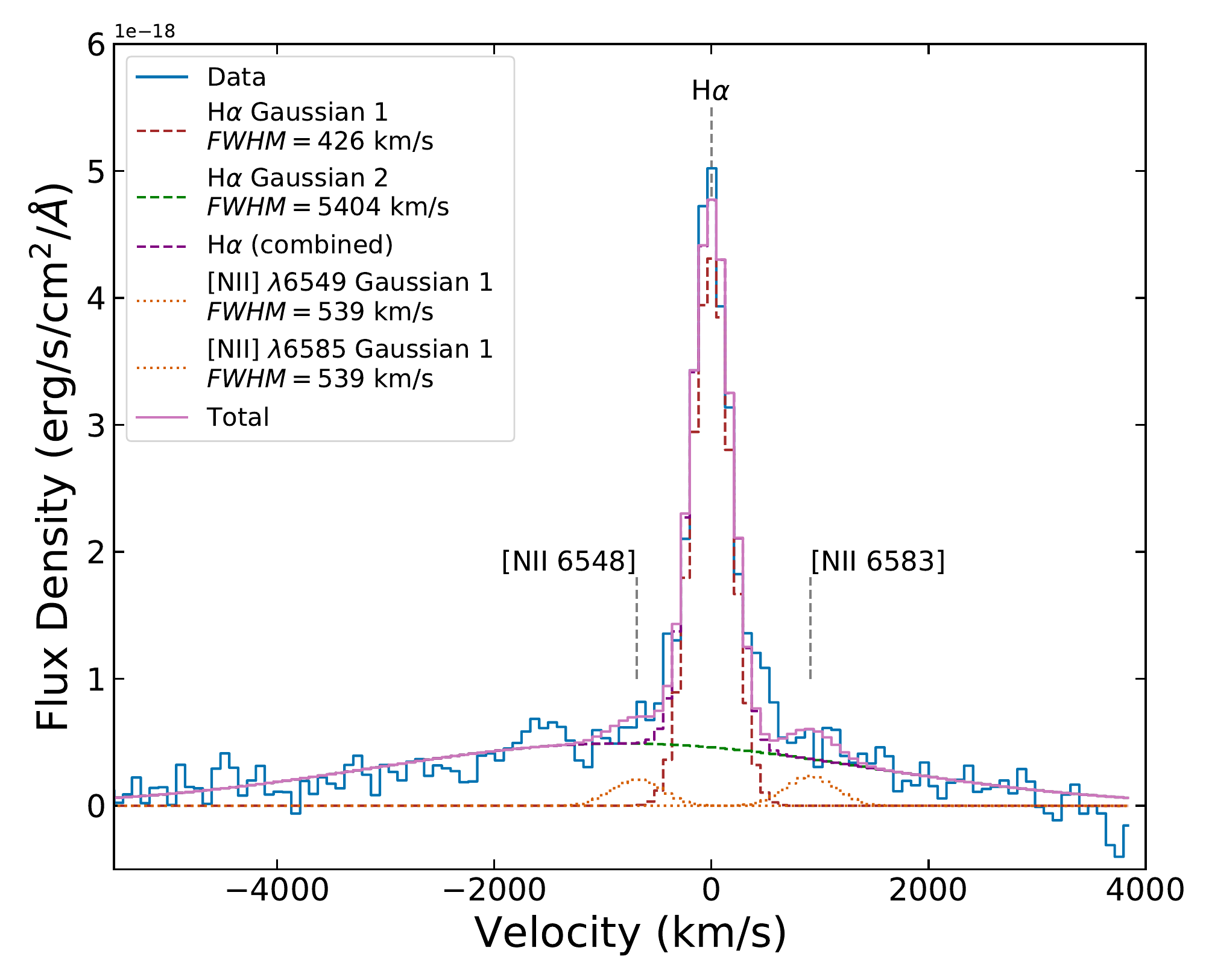}
    \label{fig:halpha}
\end{subfigure}
        
\caption{Fits to the Balmer lines H$\beta$ (left) and H$\alpha$ (right) in the +321\,d spectrum. We also show the locations of [\ion{O}{III}] and [\ion{N}{II}]. The best fit [\ion{O}{III}] fluxes are consistent with being caused by noise, while the [\ion{N}{II}] lines are marginally significant.}
\label{fig:balmer_fits}
\end{figure*}

\section{Observations}
\label{sec:obs}

AT2021lwx was discovered as a transient in Zwicky Transient Facility (ZTF; \citealt{Bellm2019,Graham2019}) imaging on the 13th of April 2021. Forced photometry revealed it to have been brightening since at least 16th June 2020 as ZTF20abrbeie. It was independently detected by the Asteroid Terrestrial-impact Last Alert System (ATLAS; \citealt{Tonry2018,Smith2020b}) as ATLAS20bkdj and the Panoramic Survey Telescope and Rapid Response System (Pan-STARRS; \citealt{Wainscoat2016,Chambers2016}) as PS22iin. AT2021lwx was twice observed by the extended Public ESO Spectroscopic Survey for Transient Objects `plus' (ePESSTO+; \citealt{Smartt2015}) and was classified as an AGN at $z=0.995$ \citep{Grayling2022}. In this section we describe photometric observations of AT2021lwx at X-ray, ultraviolet, optical, near- and mid-infrared wavelengths, which are summarised in Table \ref{tab:example} and Figure \ref{fig:lc_main}. Times $t$ are given in the rest frame at $z=0.9945$ with respect to maximum light, which we take as the date of the brightest $r$-band observation: MJD 59291.

\subsection{Optical photometry}
\label{subsec:phot_optical}
The optical light curve is compiled using photometry from publicly available ZTF $g$- and $r$-band photometry\footnote{https://lasair-ztf.lsst.ac.uk/object/ZTF20abrbeie/}, which we obtained via the ZTF forced photometry service \citep{Masci2019}. In addition, we obtained ATLAS $c$ (``cyan", $4200-6500~$\AA) and $o$ (``orange", $5600 - 8200~$\AA) imaging over the course of the light curve. The region was also covered infrequently as part of the ongoing Pan-STARRS Near Earth Object search which provides $grizy$ coverage. Data from both ATLAS and Pan-STARRS were obtained as forced photometry\footnote{https://fallingstar-data.com/forcedphot/} \citep{Smith2020b}.

To search for a host galaxy, we use the stacked $grizy$ images from the Pan-STARRS $3\pi$ survey \citep{Waters2020}. We use the \texttt{photutils} package and perform photometry in a circular aperture with a 1\,\arcsec\, radius centered on the location of AT2021lwx, and use the PS1 weight image to estimate the photometric uncertainties.

\subsection{Mid-infrared photometry}
The location of AT2021lwx was observed by the Wide-field Infrared Survey Explorer (WISE) spacecraft as part of the NEOWISE reactivation mission \citep{Mainzer2011}. We obtained NEOWISE photometry in the $W1$ and $W2$ bands from the NASA/IPAC infrared science archive\footnote{https://irsa.ipac.caltech.edu/Missions/wise.html}. Observations of the location exist with a six month cadence from several years before the flare until November 2021, during its decline. In particular we highlight MJDs 58993 (pre-flare), 59156 (on the rise), 59357 (around peak), and 59520, 59720, 59880 (during the decline). For each epoch, we use the IRSA WISE/NEOWISE coadder \citep{Masci2009} to combine individual frames. We use a sigma-clipped median to estimate the background, which we subtract and perform aperture photometry in a 4\,\arcsec\, radius around the transient location, corresponding roughly to the point spread function of WISE. Before the optical transient, we find no significant detection; after the optical flare begins there is a clear point source in both $W1$ and $W2$.

\subsection{X-ray and ultraviolet observations}
We obtained two epochs of observations with the Neil Gehrels Swift Observatory (\textit{Swift}; \citealt{Gehrels2004}). The first observation took place at MJD 59923.554 ($t_{\mathrm{max}}+317$\,d), and lasted 2670\,s. The second took place at MJD 59966.483 ($t_{\mathrm{max}}+338$\,d) and lasted 2605\,s. Data were observed in photon counting mode with the X-ray Telescope (XRT; \citealt{Burrows2005}), and in the $uvw2$ (central wavelength, 1928\,\AA), $uvm2$ (2246\,\AA), $uvw1$ (2600\,\AA), $u$ (3465\,\AA), $b$ (4392\,\AA) and $v$ (5468\,\AA) filters with the Ultraviolet-Optical Telescope (UVOT; \citealt{Roming2005}), respectively. 

The XRT data were reduced with the tasks {\sc xrtpipeline} and {\sc xselect}. The source and background events were extracted using a circular region of 40\arcsec~and an annular ring with inner and outer radii of 60\arcsec~and 110\arcsec, respectively, both centered at the position of the source. The Ancillary Response Files (ARFs) were created with the task {\sc xrtmkarf} and the Response Matrix File (RMF), swxpc0to12s6\_20130101v014.rmf, was taken from the Calibration Data base\footnote{\url{https://heasarc.gsfc.nasa.gov/docs/heasarc/caldb/swift}}.
Due to the low count rates, the XRT spectra were grouped to have a minimum of 3 counts per bin using the FTOOL {\sc grppha}.

In terms of the UVOT data, we used the task {\sc uvotimsum} to sum all the exposures when more than one snapshot was included in each individual filter data and the task {\sc uvotsource} to extract magnitudes from aperture photometry. A circular region of 5\arcsec~centered at the target position was chosen for the source event and another region of 40\arcsec~located at a nearby position was used to estimate the background emission. 
 \\
\subsection{Optical spectroscopy}
Optical spectra were obtained on MJD 59345 and MJD 59844 ($t_{\mathrm{max}}+27$\,d and $t_{\mathrm{max}}+277$\,d respectively) via the European Southern Observatory (ESO) as part of ePESSTO+ using the ESO Faint Object Spectrograph and Camera (EFOSC2; \citealt{Buzzoni1984}) on the New Technology Telescope (NTT) at ESO La Silla observatory, Chile. The first epoch consisted of a single 600\,s exposure, while the second epoch consisted of two 2700\,s exposures. All three spectra made use of grism13 ($3685-9315\,$\AA, resolution $\sim17\,$\AA). These spectra were reduced using the PESSTO pipeline \citep{Smartt2015} v3.0.1.

\subsection{Near infra-red spectroscopy}
We observed AT2021lwx on MJD 59931 and MJD 59949 ($t_{\mathrm{max}}+320$\,d and $t_{\mathrm{max}}+329$\,d respectively) with the Espectr\'ografo Multiobjeto Infra-Rojo (EMIR, \citealt{Balcells2000}) 
on the 10.8\,m Gran Telescopio Canarias (GTC) at Observatorio del Roque de los Muchachos, La Palma, Spain. Spectra were obtained under Director's Discretionary Time proposal GTC05-22BDDT (PI: M\"uller Bravo) in long-slit mode. The first epoch made use of the $YJ$ grism ($0.85 - 1.35\,\mu\mathrm{m}$, resolution $\sim10\,$\AA); the second epoch used the $HK$ grism ($1.45 - 2.42\,\mu\mathrm{m}$, resolution $\sim 20$\AA). 

Both epochs used a 1.0" slit. Image rectification, dark and flat fielding, sky subtraction, dither combination, and wavelength calibration were performed using \texttt{PyEMIR} \citep{Pascual2010,Cardiel2019} v0.17. Cosmic rays were detected and removed using the \texttt{lacosmic} algorithm \citep{VanDokkum2001}. Trace extraction was performed manually using \texttt{python}. Telluric features were removed by observing and subtracting the spectrum of the telluric standard star HIP104599. 

\section{Photometric properties}
\label{sec:phot}
The light curve of AT2021lwx (Fig. \ref{fig:lc_main}) peaked in the observer-frame $r$-band on MJD 59291, 122\,d after and 3\,mag brighter than the first ZTF detection. The decay tends to a smooth decline of $\sim 0.004$\,mag d$^{-1}$, consistent with an exponential ($e^{-0.005t}$) decay, while the late time decline is also consistent with the power-law $t^{-5/3}$. To measure the (pseudo-) bolometric light curve we $K$-correct the observed magnitudes according to \citet{Hogg2002} by assuming a black body as the underlying spectral energy distribution (SED). By correcting the observed $r$-band to the rest-frame $u$ we are able to compare to similar events at lower redshift. For those objects for which rest-frame light curves are not public, we estimate a $K-$correction to the $u$-band by choosing the observer frame band that is closest to $u$ when shifted to the rest frame. We then add the correction of $-\log(1+z)$ to account for the shortening of the wavelength range of the emission in the rest frame. The rest-frame UV peak absolute magnitude is $M \sim -25.7$\,mag. 
\subsection{Comparison to similar events}
We build a comparison sample by selecting the most luminous examples of transients from a variety of observational classes, the rest frame $u$-band absolute magnitude light curves of which we show in the lower right panel of Fig. \ref{fig:lc_main}. Our choice of objects serves two purposes: first, we search the literature for the brightest known objects of various classes; second, we show typical examples of TDEs, the most likely candidate for the type of event that could produce such extreme luminosity.

The light curve and luminosity of AT2021lwx most closely resemble a handful of transients detected in known narrow-line Seyfert 1 (NLSy1) galaxies presented by \citet{Frederick2021}, from which we select the two that display the brightest luminosity and longest timescales, AT2019brs and AT2019fdr. To these we add the even more luminous PS1-10adi \citep{Kankare2017}. NLSy1s are highly accreting AGN, and the flares have been attributed to sudden enhancements of accretion rate (either due to a TDE, or as increased gas flow) onto already-accreting SMBHs. At peak, AT2021lwx is nearly 2.5 times more luminous than any of those events.
AT2021lwx is several orders of magnitude brighter than any known supernova (SN; \citealt{DeCia2018,Angus2019,Nicholl2020a}). We show the most luminous, SN2016aps \citep{Nicholl2020a}, which has been interpreted as a possible pulsational pair instability SN (PPISN). AT2021lwx is ten times more luminous than the extreme TDE ASASSN15-lh \citep{Dong2016,Leloudas2016} and three times brighter (in the optical/UV) than the jetted TDE AT2022cmc \citep[e.g.][]{Andreoni2022,Pasham2023}. We also show three further, more regular TDEs (although the class is diverse; \citealt{Leloudas2019,VanVelzen2021}): AT2019dsg which likely includes a relativistic jet \citep{VanVelzen2021}, the well-studied AT2018dyb \citep[e.g.][]{Leloudas2019}, and the MIR-brightening PS16dtm \citep[e.g.][]{Petrushevska2023}. AT2021lwx is 100 times brighter and decays slower than these TDEs.

\subsection{Spectral energy distribution}
\label{subsec:SED}
Using the ($K$-corrected) UV photometry from $t=317$\,d we fit the spectral energy distribution (SED) with a black body with a Levenberg-Marquardt least squares minimisation. We omit the farthest UV band, uvw2, since there is evidence in the spectrum that a strong Lyman-$\alpha$ absorber is present. There is no ZTF or ATLAS photometry available at the epoch of the Swift observations, and we do not extrapolate in order to add them to the black body fit. The fit is shown in Fig. \ref{fig:SED_all}. We measure a black body colour temperature of $T_{\mathrm{C}}=1.2 \pm 0.1 \times10^4$\,K and radius of $R_{\mathrm{BB}}=1.5 \pm 0.3\times 10^{16}$\,cm, while the $t=338$\,d photometry is best fit by consistent values of ($T_{\mathrm{C}}=1.3 \pm 0.08\times10^4$\,K) and ($R_{\mathrm{BB}}=1.3\pm 0.2\times 10^{16}$\,cm).
By integrating the black body spectrum, normalised to the $r$-band peak brightness, we estimate the maximum pseudo-bolometric luminosity of the transient to be $7\times10^{45}$\,erg\,s$^{-1}$, which is similar to the median luminosity of quasars \citep{Rakshit2020}. We estimate the total energy radiated by integrating the black body with the simplistic assumption that the SED did not evolve over the duration of the transient. Up until December 2022 (440 restframe days from the onset of the event), the transient has radiated $1.5\times10^{53}$\,erg. Such a high energy release in a relatively short time is unprecedented for a transient and is usually only associated with continuously accreting SMBHs. Given that the transient becomes redder with time, this assumption likely leads to an underestimate of the peak and total bolometric luminosity.

The XRT observations resulted in the tentative detection of X-rays in the range 0.3$-$10\,keV. The second epoch is more significant at $4.6\,\sigma$, while the combined significance is $2.7\,\sigma$. Modelling the X-ray spectrum from the combined observations as a power-law, we measure an observed 0.3$-$10\,keV flux of $2.94 \pm 1.59 \times 10^{-13}$\,erg\,s$^{-1}$\,cm$^{-2}$. Correcting for Galactic neutral hydrogen column density $10^{21}$\,cm$^{-2}$ \citep{Collaboration2016} this flux corresponds to an unabsorbed luminosity of $1.52\times10^{45}$\,erg\,s$^{-1}$. This luminosity is far higher than expected by extrapolating the UV-optical black body and indicates the presence of a separate emission region. The size of the SMBH inferred from the UV-optical properties precludes the X-rays originating as thermal disk emission, while this X-ray luminosity is of similar order to the UV-optical luminosity at the same epoch. An $L_{\mathrm{optical}}/L_{\mathrm{X-ray}}\sim 1$ is consistent with late-time observations of TDEs \citep[e.g.][]{Gezari2017,VanVelzen2019,Wevers2019}.  

Emission in the MIR is also not consistent with the UV-optical black body. It continues rising after the UV-optical starts to decline, indicative either of reprocessing by dust or from a relativistic jet. Similar effects have been seen in TDEs \citep[e.g.][]{VanVelzen2016,Jiang2021,Onori2022,Petrushevska2023} and are interpreted as the echoes of the UV emission from a dusty `torus' surrounding the SMBH. From a pseudo-bolometric luminosity of $7\times10^{45}$\,erg\,s$^{-1}$ we follow \citet{Jiang2021} to estimate a sublimation radius of $\sim400$ light days, while following the $K$-band size--luminosity relation from \citet{GravityCollaboration2020} we predict $r_{\mathrm{sub}}\sim220 \pm 24$ light days. This distance should be reflected in a lag between the UV-optical and MIR emission. Although the cadence is low, we estimate the MIR to have peaked around $+110$\,d, consistent with the intrinsic scatter of the size--luminosity relation. We fit the MIR SED at the latest epoch, which is closest to the \textit{Swift} photometry and GTC spectroscopy. Fitting a black body and including the GTC $K_{S}$-band measurement, we measure a colour temperature of $T_C\simeq 10^3$\,K and a radius of $R_{\mathrm{BB}}\simeq10^{18}$\,cm, again consistent with the time lags and sublimation radius predictions.

\subsection{Host galaxy limits}
From the aperture photometry outlined in Section \ref{subsec:phot_optical} we find no significant detections in any band, leading to magnitude limits $g>23.3$, $r>23.2$, $i>23.1$, $z>22.3$, $y>21.4$. The $i$-band limit corresponds roughly to a rest-frame $g$-band absolute magnitude limit of $M\gtrsim-21.5$~mag. Assuming a mass-to-light ratio $M/L=2$ results in an stellar mass upper limit of $7\times10^{10}~M_{\odot}$. That is, an average Milky Way-like galaxy would be marginally detected in the imaging. Galaxy stellar mass ($M_*$) is correlated with the mass of central supermassive black holes. A rough $M_{\mathrm{BH}}$/$M_*$ fraction of 0.025 per cent \citep{Reines2015} implies a BH of mass  $1\times 10^{7}\,M_{\odot}\lesssim M_{\mathrm{BH}} \lesssim 3\times10^{8}\,M_{\odot}$ for a galaxy with the mass of our estimated upper limit. The PS1 catalogue \citep{Flewelling2020} contains an object located $\sim3"$ northwest of the location of AT2021lwx (PSO 318.4511+27.4315), with $r=19.25$. If we assume this object is at $z=0.9945$, then the physical separation is $30.8$~kpc, meaning it is unlikely to be associated with the transient. Using the PS1 magnitudes and their uncertainties we estimate a photometric redshift of the nearby object with \texttt{EAZY} \citep{Brammer2008} in its default configuration. We find a best fit photometric redshift $z_{\mathrm{phot}}=2.05\pm0.1$, which translates to a restframe UV brightness of $-23$~mag. This object is either a bright high redshift galaxy, quasar, or a Milky Way star, and is likely unrelated to AT2021lwx.

\section{Spectral properties}
\label{sec:spec}
The rest frame UV-optical spectra of AT2021lwx (Fig. \ref{fig:our_spec}) are dominated by Balmer emission lines superimposed on a blue continuum which becomes redder with time. Low-ionisation species common to active galactic nuclei (AGN) including \ion{Mg}{II} and \ion{He}{I} as well as semi-forbidden \ion{C}{III}] and \ion{C}{II}] are present, but various forbidden nebular lines such as [\ion{O}{II}] and [\ion{O}{III}] are absent. The spectra are compared to similar objects in Fig. \ref{fig:comparison_spec} and show similarities and differences to TDEs, SNe and AGN-flares. Particularly noticeable is the absence of lines indicative of Bowen flourescence, which appear to be ubiquitous in AGN flares and are also common in TDEs. H$\alpha$ and H$\beta$ are well represented by a two-component Gaussian profile (Fig. \ref{fig:balmer_fits}). We fit the two lines both simultaneously (although the fit is dominated by H$\alpha$ which lies in a much higher signal-to-noise ratio region) and independently. The broad component of H$\beta$ is only marginally detected. To ensure that the broad H$\beta$ component is taken into account, we fix it to the same width as that found from H$\alpha$ but allow the normalisation to vary freely. The fit results show the lines comprise an unresolved narrow component with full-width at half maximum (FWHM) $\simeq 430$ km\,s$^{-1}$, and a broad component with FWHM $5404$ km\,s$^{-1}$, the broad component being blueshifted by $\sim800$ km\,s$^{-1}$. The width and blueshift are largely dependent on the estimation of the continuum which is affected by the line corresponding to the end of the spectral coverage of the instrument. If the blueshift is confirmed, it indicates outflowing material.

To estimate a reddening we assume Case B recombination for narrow Balmer lines \citep{Osterbrock1989a}, with a temperature of $T=10,000\,$K and a density of $n_e = 100$\,cm$^{-3}$ and a \citet{Calzetti2000} extinction law (all standard choices for star-forming galaxies; we note that if the narrow Balmer lines are from the transient then Case B may not be appropriate), we measure a modest dust reddening of $E(B-V)=0.23$.
The lack of forbidden oxygen lines in the spectrum allows us to place a limit on the star-formation rate (SFR) of the host galaxy. Correcting the spectrum for Milky Way (MW) and host reddening, we estimate an upper limit on the [\ion{O}{II}] luminosity of $5.6\times10^{41}$\,erg\,s$^{-1}$. Using the $L (\mathrm{[\ion{O}{II}]})$ - SFR relation from \citet{Kennicutt1998}, modified according to \citet{Kewley2004}, we place an upper limit of $\mathrm{SFR}\leq 3.7\,\mathrm{M}_{\odot}$\,yr$^{-1}$. Assuming our upper limit on stellar mass of $7\times10^{10}~M_{\odot}$, the SFR - stellar mass relation for average star-forming galaxies at $z=1$ predicts $\mathrm{SFR}\sim 26\,\mathrm{M}_{\odot}$\,yr$^{-1}$ \citep{Zahid2012}. Unless it has a mass below  $\sim 10^9 M_{\odot}$, which would be highly unlikely given the inferred black hole mass, we conclude that the host of AT2021lwx is not a highly star-forming galaxy.

The rest-frame UV spectra of AT2021lwx display unresolved absorption features from singly-ionised metal species including \ion{Fe}{II}, \ion{Mg}{II} and possibly doubly-ionised \ion{Al}{III}. The uvw2 luminosity is a factor of 10 smaller than the rest of the UV-optical. In the rest frame at $z=0.9945$, uvw2 lies bluewards of Lyman-$\alpha$. We infer that there is a signifcant Lyman-$\alpha$ absorber located at the systemic redshift, indicating a large reservoir of warm gas, often observed in quasar \citep[e.g.][]{Viegas1995,Wolfe2005,Peroux2006} and gamma-ray burst \citep[e.g.][]{Kruehler2013,Wiseman2017a} spectra.

\section{Origin as an extreme accretion event}
\label{sec:disc}
AT2021lwx is a highly energetic event, arguably the most luminous optical transient ever observed. Such high luminosity limits the available mechanisms with which to explain the event, which we summarise in this section.

\subsection{Tidal disruption event}

In this section we discuss the possibility that AT2021lwx is a TDE. We begin by estimating some parameters of the system under the assumption of basic accretion physics and with many simplifications, before discussing the results of two different modelling approaches.

A TDE can only occur when the tidal radius lies outside the event horizon of the black hole. That tidal radius depends upon the density of the star being disrupted such that for a given stellar mass and radius there is a corresponding upper black hole mass limit, the Hills mass \citep{Hills1975}, above which a TDE cannot occur. For a non-rotating black hole, the Hills mass for a typical 1$M_{\odot}$ main-sequence star is $\sim 8\times10^7 M_{\odot}$.
Assuming a corresponding radiative efficiency of 0.1 \citep[e.g.][]{Marconi2004,Alexander2012,Nicholl2022}, the peak luminosity of AT2021lwx corresponds to a mass accretion rate of $\sim1.2\,M_{\odot}$\,yr$^{-1}$. This accretion rate lies at half of the Eddington limit for a black hole of mass $\sim10^8\,M_{\odot}$, which is far larger than the typical black hole mass inferred for TDEs \citep{Mockler2019,Wevers2019,VanVelzen2021} and marginally above the Hills mass for a 1$M_{\odot}$ star. Assuming an accretion rate typical of quasars ($\dot{M}/\dot{M}_{\mathrm{edd}}=0.1$) implies $M_{\mathrm{BH}}\sim10^9\,M_{\odot}$. This lies close to the mean black hole mass for quasars at $z=1$ which is $10^{8.5}\,M_{\odot}$ \citep{McLure2004}. This black hole mass is beyond the Hills mass for stars up to $200\,M_{\odot}$, unless the black hole is rotating with a large spin, in which case the Hills mass increases \citep{Leloudas2016}, or if the star is evolved which has the same effect due to the lower density of its atmosphere. At an accretion rate of $1-2\,M_{\odot}$\,yr$^{-1}$, over two solar masses must already have been accreted, placing that mass as a lower limit on any star to have been disrupted. 

The energetics of the event are compatible with a TDE if we expand the range of progenitor stars to far higher masses than have been observed before. Fitting the light curve using Modular Open Source Fitter for Transients (\textsc{MOSFiT}; \citealt{Guillochon2018}) using the TDE model of \citet{Mockler2019} we estimate $M_{\mathrm{BH}} = 8.3\times10^8\,M_{\odot}$ and $M_* = 14.8\,M_{\odot}$, consistent with accretion close to the Eddington limit. It has previously been assumed that TDEs of such massive stars are extremely rare, not least because the lifetimes of such massive stars are of the order 15\,Myr \citep{Meynet2002}. A star of this mass must therefore be born within the very central region of the galaxy in order to pass within the tidal radius of the SMBH within its lifetime, which we find unlikely. Similarly, the existence of a $\sim15\, \mathrm{M}_{\odot}$ star requires strong star formation in the host galaxy which is generally inconsistent with our upper limits. Nevertheless, the MW Galactic centre hosts a population of young ($\lesssim 10-100$\,Myr) massive ($\gtrsim 10\,M_{\odot}$) stars, despite there still not being a clear understanding of their origin \citep[e.g.][]{Paumard2006,Lu2008}. Recent
 work has also suggested that there are more tidal disruptions of higher-mass stars than previously expected – possible explanations include higher SFR in the centres of these galaxies compared to elsewhere in the galaxies or top-heavy initial mass functions in galactic nuclei producing more higher-mass stars in these regions \citep{Mockler2022}. Furthermore, the apparent overabundance of carbon compared to oxygen is consistent with some observations of evolved asymptotic giant branch stars \citep[e.g.][]{DeBeck2020}. Even so, the lower limits from this study were in the $1 - 2\,M_{\odot}$ range, which is far below the mass inferred for AT2021lwx. Furthermore, we note that there is a strong degeneracy in MOSFiT between the radiative efficiency and the stellar mass of the accreted star \citep{Mockler2021}.

 It should also be noted that MOSFiT has not been designed to model disruptions of massive stars; it currently only includes zero-age main sequence stars, while a massive star may well be evolved into a more diffuse, giant stage. Indeed, the disruption of evolved stars, or those with diffuse envelopes, is a plausible mechanism for a TDE \citep[e.g.][]{Macleod2012,Law-Smith2017}, and while uncertainty in the mass-radius relation from stellar evolution is included in MOSFiT’s systematic error measurements, an extreme example could push these errors to the maximum. 

Instead of MOSFiT, which relies on simulations of \textit{disruption} and assumes a set of standard relations to describe the transfer of energy from the disrupted material, through the accretion disk, into radiation, we attempt to model the light curve using specific \textit{disk} models that assume an initial disk condition. Following the work of \citet{Mummery2020a} we are able to explain the peak luminosity and late-time UV-optical SED of AT2021lwx with a $2\,M_{\odot}$ star being disrupted by a higher mass, maximally spinning SMBH of $M_{\mathrm{BH}} > 1\times10^9\,M_{\odot}$. These results are reliant on a compact initial disc (around the innermost stable circular orbit; ISCO), and a short viscous timescale on the order of the orbital timescale -- usually disc timescales are expected to be several orders of magnitude larger than the orbital timescale. We find that including a much larger disc mass (i.e. mass of the disrupted star) drives the temperature to be inconsistent with observations (although this could be rectified if the observations are influenced by significant host galaxy reddening). 

 Neither of the above models accounts for the possibility of super-Eddington accretion, which is likely to occur in TDEs \citep{Dai2018}, particularly in the early phases of the event. In the super-Eddington regime, the disk is likely to inflate into a geometrically thick disk at which point many of the assumptions used in those models break down. For example, there will be significant mass loss from the disk due to winds, viewing angle effects, and the likely launching of relativistic jets. 

The spectra of AT2021lwx do not resemble typical TDEs \citep[e.g.][]{Leloudas2019,VanVelzen2021,Charalampopoulos2022}. Around half of TDEs appear to show \ion{He}{II}, \ion{Fe}{II} and/or Bowen features, and while it is fairly common for TDEs to exhibit only Balmer features, such features tend not to include such strong narrow (in our case unresolved) components as AT2021lwx (although see PS16dtm; \citealt{Petrushevska2023} and AT2019dsg; \citealt{VanVelzen2021}). If the transient is caused by the disruption of a star, then these spectral features must originate from a reprocessing region not present in typical TDEs. Instead, the narrow components resemble more closely the low-velocity, shocked circumstellar material that gives rise to the narrow lines of type IIn SNe.

\subsubsection{The expected rate of high-mass star TDEs}
To assess the plausibility of detecting a TDE of a $\sim 15\, M_{\odot}$ star, we make a brief estimate of the expected rate of such events. Measurements of the volumetric and galactic rates of TDEs are sparse, but progress has been made with systematically selected samples from ZTF \citep{Yao2023} who find $3.2\times 10^{-5}$\,galaxy$^{-1}$\,yr${-1}$ in galaxies of $10^{10}\,M_{\odot}$. To estimate the intrinsic rate of TDEs of $15\,M_{\odot}$ stars we take the rate of \citet{Yao2023}, assume that it is representative for $1\, M_{\odot}$ TDEs, and scale it to $15\,M_{\odot}$ by the stellar initial mass function (IMF). Of the 33 TDE’s in their sample, only one has a stellar mass measurement substantially greater than $1\,M_{\odot}$. We approximate the rate of TDEs of different mass stars by assuming that it scales directly with the IMF. Using an IMF with high-mass slope $\alpha=-2.3$ \citep[e.g.][]{Salpeter1955, Kroupa2001,Chabrier2003}, there are 500 times fewer $15\,M_{\odot}$ stars than solar mass stars. The fraction of disruption candidates at this mass is smaller still due to the short lifetimes of these stars (a $15\,M_{\odot}$ star lives $10^3$ times shorter than a $1\,M_{\odot}$ star), the exact reduction dependent upon the integrated stellar population age of stars in the galaxy nucleus. This effect is reflected in the stellar mass function of TDEs: \citet{Nicholl2022} showed that the disrupted stellar mass function drops off faster than the IMF (although the mass function is relatively unconstrained at masses $\gtrsim 1.5\,M_{\odot}$), since there are only one or two known TDEs from stars with a mass greater than that number. This is likely due to the short lifetime of such intermediate and high mass stars. Combining the ratio of lifetimes with the IMF, the intrinsic rate of $15\,M_{\odot}$ TDEs is likely $\sim10^6$ times smaller than intrinsic rate of $1\,M_{\odot}$ TDEs, rendering the discovery unlikely. However, assuming the energetics are correct and that a TDE of a $15\,M_{\odot}$ star is roughly 100 times brighter than that of a solar mass star, then those from higher mass stars would be detectable by ZTF in a greater volume. Using the faintest magnitude limit used by \citet{Yao2023}, $15\,M_{\odot}$ TDEs would be visible in a volume 120-times larger. Thus, the observed rates of $15\,M_{\odot}$ TDEs may only be $>10^4$ times smaller than those at smaller masses. Nevertheless, with current samples numbering in the few tens of objects, it is thus highly unlikely that a $15\,M_{\odot}$ TDE would have been discovered.

Overall, while the light curve, SED and spectra can be explained by a TDE, the frequency and nature of such a disruption is highly unconstrained and significant advances in the modelling of both the disruption of massive stars, and the physics of the subsequent accretion, must be made before strong conclusions are drawn.

\subsection{Turn on of an AGN via sudden accretion of gas}
An alternative to the disruption of a star is the sudden, and isolated, accretion of a large amount of more diffuse material. This is distinct from the classical CLAGN \citep[e.g.][]{LaMassa2015,Frederick2019,Graham2020} which tend to show a change of \textit{spectral} properties. The turn-on scenario has been proposed for the smooth flaring events in known AGN such as AT2017bgt \citep{Trakhtenbrot2019} and AT2019brs \citep{Frederick2021}. Unlike those transients, there is no evidence for a pre-existing AGN in AT2021lwx, although a low-luminosity AGN is not ruled out due to the larger cosmological distance. \\

A handful of optical nuclear flares that do not adhere to typical TDE, CLAGN, or SN characteristics have been accumulated over recent years \citep{Kankare2017,Trakhtenbrot2019,Frederick2021}.
Optically, AT2021lwx is more luminous than all of these events, while the light curve shape is similar to those interpreted as unusual accretion events. Spectroscopically, the two-component H$\alpha$ profile is similar to the nuclear transient PS1-10adi, with the broad component showing a similar velocity to that transient and AT2017bgt, a long-lasting optical flare in a known AGN. The width of the broad component in AT2021lwx is larger than all other similar nuclear flares in \cite{Frederick2021}. Unlike all transients in the comparison sample, the spectrum of AT2021lwx shows no \ion{Fe}{II} or features excited by Bowen fluorescence (e.g. \ion{N}{III}). 

The large widths of the broad lines imply emission from a region with a high-velocity dispersion, while the existence of narrow lines implies a slow-moving component. Such a scenario resembles a traditional AGN according to the unification picture \citep{Antonucci1993}, where broad lines emanate from the eponymous broad-line region (BLR), comprising clouds of gas that follow virialised orbits around a supermassive black hole. If the observed broad Balmer emission originates from an illuminated virialised BLR, then the radius of this region is $r_{\mathrm{BLR}}\sim5\times10^{16} - 5\times 10^{17}$\,cm for black holes of $10^8\,M_{\odot} - 10^9\,M_{\odot}$ respectively. This inferred BLR size is similar in scale to the inferred black body radius $R_{\mathrm{BB}}$, although it is likely that the assumption of a spherical photosphere does not hold if the UV emission originates from an accretion disk.

The narrow line region (NLR) in AGN comprises gas at much larger radii, with light-travel times from black hole to NLR of $10^2 - 10^5$ yr, too far to have been ionised by the current flare of AT2021lwx. If the narrow Balmer lines in the AT2021lwx spectra correspond to a narrow line region in its host galaxy, they must have been ionised by a previously UV-bright source (i.e. an AGN's accretion disk). However, the lack of narrow nebular lines other than from hydrogen and carbon renders the NLR unlikely as the source of these features. A second possible explanation for the lack of [\ion{O}{II}], [\ion{O}{III}] and narrow \ion{Mg}{II} is the Baldwin effect \citep{Baldwin1977}, an observed anti-correlation between AGN luminosity and emission line equivalent widths. Indeed, according to the relationship between B-band absolute magnitude and [\ion{O}{II}] equivalent width in \citet{Croom2002} we may not expect to see any [\ion{O}{II}] emission at all. [\ion{O}{II}] is found to be lacking in around 5 per cent of all AGN \citep{Schawinski2015}.


\subsection{Accretion of a gas cloud by a dormant black hole}
The lack of evidence for an AGN coupled with the difficulties of explaining the event with a TDE, we explore an alternative origin of AT2021lwx. It is plausible that a large amount of material in the form of a molecular cloud was disrupted (see e.g., \citealt{Wang2017}) and accreted by the black hole. Assuming a TDE-like scenario, the fuelling rate is determined by the fallback rate of material onto the black hole after the disruption. To achieve the observed rise time, the cloud would need to be compact enough to provide significant fuel at early times, since a diffuse cloud would likely result in a slower fallback rate. Detailed modelling of this scenario is deferred to further work.

\subsection{Lensed superluminous supernova}
The smooth and slow rise, long decline, and colour evolution of AT2021lwx resemble superluminous SNe (SLSNe), which are thought either to be powered by magnetars \citep[e.g.][]{Woosley2010,Kasen2010} or interaction of ejecta with dense circumstellar material (CSM). The latter scenario is particularly likely in the case of SLSNe-IIn \citep[e.g.][]{Smith2007,Benetti2014,Nicholl2020a} which display narrow Balmer features as seen in AT2021lwx, raising the possibility of a SN origin for the transient. A number of observed properties deter us from this conclusion. Primarily, the peak luminosity requires an extremely unlikely mass of CSM, of the order $250\,M_{\odot}$ following \citet{Chevalier2011}, larger even than the zero-age main sequence mass of the largest expected SN progenitors \citep[e.g.][]{Kasen2011}. The required CSM mass is reduced if the lightcurve is strongly lensed, and thus magnified, by a foreground source. Given the lack of evidence for any galaxy foreground galaxy in deep imaging, nor any system at a lower redshift in the spectra, we find the lensing explanation unlikely. 
\section{Conclusions}

AT2021lwx is an extraordinary event that does not fit into any common class of transient. With a total radiated energy $>10^{53}$\,erg, it is one of the most luminous transients ever discovered. By collecting and analysing multiwavelength photometry and spectroscopy of the transient, we conclude that:
\begin{enumerate}
    \item[i)] the emission is dominated by a black body with a moderate temperate of 12,000\,K and a large radius of $10^{16}$\,cm which cools as the transient fades
    \item[ii)] there are two components to the material, one with a large velocity dispersion ($\sim 5400$\,km\,s$^{-1}$) and another slow-moving component ($\sim 430$\,km\,s$^{-1}$), the fast-moving component potentially forming an outflow
    \item[iii)] the luminosity is likely powered by the accretion of a large amount of gas onto a supermassive black hole with mass $>10^{8}\,\mathrm{M}_{\odot}$
    \item[iv)] while there is no evidence of AGN activity, there is, via non-thermal X-rays, evidence for a hot corona or jet as well as the presence of large amounts of dust
    \item[v)] a tidal disruption of a massive star is unlikely due to the small chance of such an event observing
    \item[vi)] the spectral and photometric features of the transient suggest the sudden accretion of a large amount of gas, potentially a giant molecular cloud.
\end{enumerate}
Further follow-up and modelling of AT2021lwx is necessary to reveal more about the scenario that caused the flare, and the community is strongly encouraged to search for similar events in both the future as well as in archival data.

\section*{Acknowledgements}

We thank the anonymous referee for their helpful comments. We thank Nick Stone for helpful discussions on the interpretation of the transient. We are grateful for the rapid turnaround of the DDT proposal by the time allocation committee of the GTC. PW acknowledges support from the Science and Technology Facilities Council (STFC) grant ST/R000506/1.
YW acknowledges support from the Royal Society Newton Fund.
M.P and G.L. are supported by a research grant (19054) from VILLUM FONDEN. T.E.M.B. and L.G. acknowledge financial support from the Spanish Ministerio de Ciencia e Innovaci\'on (MCIN), the Agencia Estatal de Investigaci\'on (AEI) 10.13039/501100011033, the European Social Fund (ESF) "Investing in your future, and the European Union Next Generation EU/PRTR funds under the PID2020-115253GA-I00 HOSTFLOWS project, the 2019 Ram\'on y Cajal program RYC2019-027683-I, the 2021 Juan de la Cierva program FJC2021-047124-I, and from Centro Superior de Investigaciones Cient\'ificas (CSIC) under the PIE project 20215AT016, and the program Unidad de Excelencia Mar\'ia de Maeztu CEX2020-001058-M. MN is supported by the European Research Council (ERC) under the European Union’s Horizon 2020 research and innovation programme (grant agreement No.~948381) and by funding from the UK Space Agency. P.C. acknowledges support via an Academy of Finland grant (340613; P.I. R. Kotak). This work was partially funded by ANID, Millennium Science Initiative, ICN12\_009. MG is supported by the EU Horizon 2020 research and innovation programme under grant agreement No 101004719. NI was partially supported by Polish NCN DAINA grant No. 2017/27/L/ST9/03221. This work was supported by a Leverhulme Trust International Professorship grant [number LIP-202-014]. TP acknowledges the financial support from the Slovenian Research Agency (grant P1-0031). FV is supported from the Spanish Ministry of Science and Innovation research project PID2020-120323GB-I00 and from grant FJC2020-043334-I financed by MCIN/AEI/10.13039/501100011033 and NextGenerationEU/PRTR.

Based on observations collected at the European Organisation for Astronomical Research in the Southern Hemisphere, Chile, as part of ePESSTO+ (the advanced Public ESO Spectroscopic Survey for Transient Objects Survey).
ePESSTO+ observations were obtained under ESO program IDs 1103.D-0328, 106.216C and 108.220C (PI: Inserra). Based on observations made with the GTC telescope, in the Spanish Observatorio del Roque de los Muchachos of the Instituto de Astrofísica de Canarias, under Director’s Discretionary Time GTC05-22BDDT (PI: M\"uller
Bravo). 
This work made use of \textsc{photutils} \citep{Bradley2022}.

\section*{Data Availability}

\textit{Swift} X-ray and UV data are publicly available from the UK Swift Science Data Centre, \url{https://www.swift.ac.uk/swift_portal}. ZTF data can be retrieved from the ZTF forced photometry service, \url{https://web.ipac.caltech.edu/staff/fmasci/ztf/forcedphot.pdf}. The GTC spectra will be made public upon acceptance via the GTC archive. WISE data are publicly accessible via the NASA IPAC infra-red science archive (IRSA) at \url{https://irsa.ipac.caltech.edu}.



\bibliographystyle{mnras}
\bibliography{PhilMendeley} 




%
%


\bsp	
\label{lastpage}
\end{document}